\documentstyle[11pt]{article}

\font\tenrm=cmr10
\font\tenit=cmti10
\font\elevenbf=cmbx10 scaled\magstep 1
\font\elevenrm=cmr10 scaled\magstep 1
\font\elevenit=cmti10 scaled\magstep 1

\font\ninerm=cmr9

\renewenvironment{thebibliography}[1]
 { \elevenrm
   \begin{list}{\arabic{enumi}.}
    {\usecounter{enumi} \setlength{\parsep}{0pt}
     \setlength{\itemsep}{3pt} \settowidth{\labelwidth}{#1.}
     \sloppy
    }}{\end{list}}

\begin{document}
\begin{flushright}
{\elevenbf TTP93-21\\
July 1993}
\end{flushright}
\begin{center}
\vglue 0.6cm
{ {\elevenbf        \vglue 10pt
               RADIATIVE CORRECTIONS TO THE TOP QUARK WIDTH\footnote{
\ninerm
\baselineskip=11pt
Talk presented by M. Je\.zabek at the Workshop on Physics
and Experiments at Linear $e^+e^-$ Colliders,
Waikoloa, Hawaii, April 1993} \footnote{Work partly
supported by the Polish
Committee for Scientific
Research (KBN) Grants 203809101 and 223729102, and
by EEC Contract ERB-CIPA-CT-92-2077}
\\}
\vglue 1.0cm
{\tenrm Marek JE\.ZABEK\\}
\baselineskip=13pt
{\tenit Institute of Nuclear Physics, Kawiory 26a, PL-30055 Cracow,
Poland\\}
\baselineskip=12pt
{\tenrm and\\}}
\baselineskip=12pt
{\tenit Institut f. Theoretische Teilchenphysik, Universit\"at
Karlsruhe\\}
{\tenit 76128 Karlsruhe, Germany\\}

\vglue 0.3cm
{\tenrm and\\}
\vglue 0.3cm
{\tenrm Johann H. K\"UHN and Thomas TEUBNER\\}
{\tenit Institut f. Theoretische Teilchenphysik, Universit\"at
Karlsruhe\\}
{\tenit 76128 Karlsruhe, Germany\\}
\vglue 0.8cm
{\tenrm ABSTRACT}

\end{center}

\vglue 0.3cm
{\rightskip=3pc
\leftskip=3pc
\tenrm\baselineskip=12pt
\noindent
Calculations of radiative corrections to the top quark width are
reviewed. QCD effects are discussed for $t-\bar t$ systems
produced in $e^+e^-$ annihilation near the energy threshold.}
\vglue 0.6cm
{\elevenbf\noindent 1. Single Quark Decay}
\vglue 0.2cm
{\elevenit\noindent 1.1. Total Rate}
\vglue 0.1cm
\baselineskip=14pt
\elevenrm
The top is the first heavy quark whose mass can
be measured to better than 1\% precision at a future $e^+e^-$
collider \cite{Workshop,ExA,Martinez}~.
Therefore, measurements of its width will not only test
the standard model at the Born level, but
also the QCD radiative corrections
which are of order 10\% \cite{JK1}~.
This is in contrast to
$b$ and $c$ quarks, where uncertainties in
the masses and non-perturbative effects preclude
this possibility.\par
The one loop electroweak corrections to
the total rate have been also calculated.
These corrections evaluated in the narrow
width approximation \cite{DS,Gad}
turned out to be positive and rather small (1-2\%).
A reason for that is that the contribution from the
Higgs field Yukawa coupling remains very small for
realistic top quark masses \cite{IMT}~.
The effect of finite $W$
width \cite{Topw} is comparable in size
to the narrow width electroweak correction but of the
opposite sign for $m_t$ above $110$ GeV.\par
Formulae for the QCD and electroweak  corrections
to the top quark width in the standard model are given in
Ref.\cite{Topw} as well as a more comprehensive list of
papers on this subject. In Table 1 we summarize
the results \cite{Topw} for the corrections obtained from
different approximations as well as for the total decay rate
$\Gamma_t$ and its narrow width Born approximation
$\Gamma^{(0)}_{nw}$~.
We give the ratios of the corrections to the
zeroth-order result       $\Gamma^{(0)}_{nw}$~,
i.e. we define
\begin{equation}
\delta^{(i)} = \Gamma^{(i)}/\Gamma^{(0)}_{nw} - 1
\label{eq:13}
\end{equation}
where $i = 0,1$
corresponds to the Born and the QCD corrected rate
respectively, and the widths in the numerators include the
effects of the W propagator.  Analogously we define
$\delta^{(1)}_{nw}$
which is given by the ratio of the QCD corrected and the Born
widths, both evaluated in the narrow width approximation,
$\delta^{(1)}_{nw}$$(0)$
for massless $b$ quark and
$\delta_{ew}$ for the electroweak narrow width result~\cite{DS}~.

\begin{table}[h]
\begin{tabular}{|r|c|c|c|c|c|c|c|c|c|} \hline
$m_t\ $  & $\alpha_s(m_{t})$ & $\Gamma^{(0)}_{nw}$ &$\delta^{(0)}_{}$&
$\delta^{(1)}_{nw}$$(0)$ & $\delta^{(1)}_{nw}$ &
$\delta^{(1)}_{}$&$\Gamma^{(1)}_{}$&$\delta^{}_{ew}$&$\Gamma^{}_{t}$\\
{\scriptsize(GeV)} &  & {\scriptsize(GeV)} & {\scriptsize(\%)} &
{\scriptsize(\%)} & {\scriptsize(\%)} & {\scriptsize(\%)} &
{\scriptsize(GeV)} & {\scriptsize(\%)} & {\scriptsize(GeV)} \\
\hline
  90.0& .118& .0234& 11.69 & 7.88 &-3.81&  6.56 &.0249& 0.81& .0251\\
 100.0& .116& .0931&  0.16 &-4.56 &-6.91& -6.89 &.0867& 1.04& .0876\\
 110.0& .115& .1955& -1.44 &-6.81 &-7.83& -9.22 &.1775& 1.20& .1796\\
 120.0& .113& .3265& -1.78 &-7.61 &-8.20& -9.89 &.2942& 1.33& .2982\\
 130.0& .112& .4849& -1.82 &-7.97 &-8.37&-10.08 &.4360& 1.43& .4423\\
 140.0& .111& .6708& -1.77 &-8.15 &-8.44&-10.10 &.6031& 1.51& .6122\\
 150.0& .110& .8852& -1.69 &-8.25 &-8.47&-10.05 &.7962& 1.57& .8087\\
 160.0& .109& 1.130& -1.60 &-8.31 &-8.49& -9.99 &1.017& 1.62& 1.033\\
 170.0& .108& 1.405& -1.52 &-8.34 &-8.49& -9.91 &1.266& 1.67& 1.287\\
 180.0& .107& 1.714& -1.45 &-8.35 &-8.48& -9.84 &1.546& 1.70& 1.572\\
 190.0& .106& 2.059& -1.39 &-8.36 &-8.47& -9.77 &1.857& 1.73& 1.890\\
 200.0& .106& 2.440& -1.33 &-8.36 &-8.46& -9.70 &2.203& 1.76& 2.242\\
\hline
\end{tabular}
\caption{{\tenrm
Top width as a function of top mass and the comparison of
the different approximations.}}
\end{table}

A number of intrinsic uncertainties remains.
It should be noted that the size of the electroweak corrections
is comparable to the uncertainties from as yet uncalculated
${\cal O}({\alpha_s}^2)$ correction.
The present
uncertainty in $\alpha_s$ and the ignorance concerning the
the second order QCD correction
limit the accuracy of the
prediction to about 1-2\%.
One has to take into account the
experimental and theoretical~\footnote{\ninerm M.J. thanks
Andrzej Buras for a helpful discussion on this subject}
errors
in the determination of the top mass which may
lead to uncertainties of similar magnitude, in particular
for lower allowed values of $m_t$~.
At present the best place for a precise determination of
$\Gamma_t$ is believed to be the threshold region for
$t\bar t$ production in $e^+e^-$ annihilation. The most
optimistic current estimate of the relative precision is
5\% \cite{Fujii}~, so at present the theory seems to be
in good shape. However, in the future when $e^+e^-\ 500$
becomes a reality
it will be mandatory to give the theory prediction
including the ${\cal O}({\alpha_s}^2)$ contribution.
Bound state effects in the threshold region, c.f. next
section, may in principle enhance this correction.
Needless to say such a calculalation is necessary
when one aims, as many people do, to use a precise
measurement of the top width as a consistency check
of the standard model.\par
In fact a number of calculations have been performed studying
electroweak effects on the top width in theories extending
the standard model. In particular it has been
found~\cite{GH} that the additional corrections
from the extended Higgs sector of the minimal supersymmetric
standard model are significantly smaller than 1\%~.
The situation changes drastically when the chanel $t\to H^+b$
is kinematically allowed. QCD corrections to the corresponding
partial width have been recently calculated \cite{Yuan}
as well as the electroweak ones \cite{Czarnecki}~.
\vglue 0.2cm
{\elevenit\noindent 1.2. Differential Distributions}
\vglue 0.1cm
The calculations of QCD corrections to the differential
decay distributions have been reviewed in \cite{ThA}~.
Recently a calculation \cite{CJK} of the $W$ mass distribution
in $t\to b\bar ff^\prime$ including ${\cal O}({\alpha_s})$
corrections
has been repeated and a fast Monte Carlo generator for these
decays has been written \cite{Juenger}~.
\vglue 0.6cm
{\elevenbf\noindent 2. Width of $t-\bar t$ system near threshold}
\vglue 0.2cm
{\elevenit\noindent 2.1. Motivation}
\vglue 0.1cm
\baselineskip=14pt
\elevenrm
{}From Table 1 and the present Fermilab lower limits on
the top quark mass we conclude
that the $t$ quark is a short--lived particle, and its width
$\Gamma_t$ is of the order of several hundred MeV. As a
consequence the cross section for $t\bar t$ pair production
near energy threshold has a rather simple and smooth shape.
In particular, it is likely that in $e^+e^-$ annihilation
only the $1S$ peak survives as a remnant of toponium resonances.
Nevertheless,
the excitation curve $\sigma(e^+e^-\rightarrow t\bar t\,)$
allows a precise determination of $m_t$ and the strong coupling
constant $\alpha_s$~\cite{FK,SP,ExA}~.
The idea \cite{FK,SP} to use the Green function
instead of summing over overlapping resonances has been also
applied in calculations of differential cross sections,
in particular for intrinsic momentum distributions of top quarks
in $t\bar t$ systems \cite{Sumino,JKT}~. It has been argued
\cite{Sumino1} and demonstrated \cite{Martinez} that
the combined measurements of the total
and the differential cross sections in
$e^+ e^- \to t\bar t$
offer a very promissing method for a simultaneous
determination of $m_t$ and $\alpha_s$~.
A possible problem is related to the fact that when produced
near energy threshold $t$ and $\bar t$ cannot be considered
as free particles. The binding energy and the `intrinsic'
kinetic energy  of the $t-\bar t$ system
tend to reduce the available
phase space for the decay. Although the effect is only
${\cal O}({\alpha_s}^2)$ the suppression is large \cite{JK}~,
especially for $m_t$  slightly above the threshold for real
$W$ decay. Apparently in a high precision calculation
one has to consider the width $\Gamma_{t-\bar t}(p)$ as
a non-trivial function of the intrinsic momentum $p$~.
It is not surprising at all that when the phase space
suppression is taken into account one finds \cite{Sumino}
that the effects of the momentum dependent width
are quite large and may show up in the annihilation cross section
$\sigma(e^+e^-\rightarrow t\bar t\,)$~.
An immediate question is: to what extend do theoretical
model assumptions spoil the precision of determination
of $m_t$ and $\alpha_s$~?
\vglue 0.2cm
{\elevenit\noindent 2.1. Models and Results}
\vglue 0.1cm
\baselineskip=14pt
The width of the $t-\bar t$ system depends on
the intrinsic momentum, of say $t$ quark, because both the
matrix element and the phase space available for the decay products
depend on it. The phase space effect tends to reduce the decay rate
of bound top quarks relative to free ones \cite{JK}, and the
effect is enhanced, because for short--lived
particles the distribution of intrinsic momentum is broad.
However, for the same reason the decays
take place at short relative distances, where the wave functions
of $b$ and $\bar b$ quarks originating from the decays are
distorted ({\em enhanced}) by Coulomb attraction. Therefore, when
calculating the amplitude of $t\rightarrow bW$ transition,
one should use Coulomb wave functions rather than plane waves
for $b$ quarks. This effect clearly increases the rate. A third
factor is the time dilatation: a top quark moving with velocity
$v$ lives longer in the center--of--mass laboratory frame.
While phase space reduction and time dilatation can be
implemented in a straightforward way Coulomb enhancement
cannot be easily taken into account. In principle
one has to replace the plane wave functions for $b$ quarks by
relativistic Coulomb functions when evaluating the amplitude
for the $t\rightarrow bW$ transition. One may hope, however, that
the following observation, valid for muons bound in nuclei
\cite{Huff}, holds also for chromostatic attraction in
$t-\bar t$ systems: the phase space suppression and the Coulomb
enhancement nearly cancel each other. For light nuclei the result
is well described by the time dilatation suppression alone\footnote{
\ninerm\baselineskip=11pt
For $\mu^-$ bound in a nucleus
of charge $Z$ one obtains
$$\Gamma = \Gamma_{free}
\left[1-5(Z\alpha)^2\right]\left[1+5(Z\alpha)^2\right]\left[
1-(Z\alpha)^2/2\right]$$ where the first correction factor
comes from the phase space suppression, the second from the Coulomb
enhancement, and the third one from time dilatation. Thus there is
no first order correction to the total rate from the rescattering
in the nucleus potential \cite{Huff}~.
A similar result has been recently obtained for the final
state rescattering in $t-\bar t$ threshold region
\cite{Sumino2}~.}.
Using this observation we show  \cite{JT}
that despite the phase space suppression
the effect of the $p$-dependent width on both the total
$\sigma(e^+e^-\rightarrow t\bar t\,)$ and the differential
${{\rm d}\sigma/{\rm d}p}$ cross section
is likely to be small.
Moreover, we show that even if the effects of the momentum
dependent width were important the resulting uncertanties
for $\alpha_s$ and $m_t$ would be reasonably small.\par
We show our results for $m_t$~=~120~GeV. This is likely to be
the most difficult case. For higher masses the effects of the
momentum dependent width are smaller. For lower $m_t$ more
information is available from peaks in the total cross section.
We compare the total $\sigma(e^+e^-\rightarrow t\bar t\,)$
(Figure 1) and differential (Figure 2)
${{\rm d}\sigma/{\rm d}p}$
cross sections. The dashed lines are obtained assuming constant
momentum independent width.
The dotted lines correspond to Model 1
where the momentum dependent width
$\Gamma_{t-\bar t}(p)$  is significantly reduced for intermediate
and large momenta $p$, mainly as a consequence of phase space
suppression.
The solid lines (Model 2) have been obtained assuming cancelation
of phase space suppression and Coulomb enhancement.
It can be seen that the results of this model are quite close
to those obtained assuming
constant width. It is noteworthy that even
for Model 1 the $1S$ peak (threshold position) in
$\sigma(e^+e^-\rightarrow t\bar t\,)$
and the position of the
maximum for ${{\rm d}\sigma/{\rm d}p}$
are not much affected by the momentum dependent width.
Since the idea of \cite{Martinez} is to combine just these
observables the resulting theoretical errors in determination
of $m_t$ and $\alpha_s$ are quite small.\par
%
%
\newpage
{\elevenbf\noindent  References \hfil}
\vglue 0.4cm

\end{document}